\documentclass[twocolumn,superscriptaddress]{revtex4}
\usepackage{amsmath}
\usepackage{amsfonts}
\usepackage{braket}
\usepackage{graphicx}

\begin{document}

\title{Quantum simulations with multiphoton Fock states}

\author{T.~Sturges}
\thanks{These two authors contributed equally}
\affiliation{Institute of Theoretical Physics, University of Warsaw, ul.\ Pasteura 5, 02-093 Warsaw, Poland}
\author{T.~McDermott}
\thanks{These two authors contributed equally}
\affiliation{Institute of Theoretical Physics, University of Warsaw, ul.\ Pasteura 5, 02-093 Warsaw, Poland}
\author{A.~Buraczewski}
\affiliation{Institute of Theoretical Physics, University of Warsaw, ul.\ Pasteura 5, 02-093 Warsaw, Poland}
\author{W.~R.~Clements}
\affiliation{Clarendon Laboratory, University of Oxford, Parks Road, Oxford OX1 3PU, United Kingdom}
\author{J.~J.~Renema}
\affiliation{Complex Photonic Systems (COPS), MESA+ Institute for Nanotechnology, University of Twente, P.O. Box 217, 7500 AE Enschede, Netherlands}
\author{S.~W.~Nam}
\affiliation{National Institute of Standards and Technology, 325 Broadway, Boulder, CO 80305, USA}
\author{T.~Gerrits}
\affiliation{National Institute of Standards and Technology, 325 Broadway, Boulder, CO 80305, USA}
\author{A.~Lita}
\affiliation{National Institute of Standards and Technology, 325 Broadway, Boulder, CO 80305, USA}
\author{W.~S.~Kolthammer}
\affiliation{Clarendon Laboratory, University of Oxford, Parks Road, Oxford OX1 3PU, United Kingdom}
\author{A.~Eckstein}
\affiliation{Clarendon Laboratory, University of Oxford, Parks Road, Oxford OX1 3PU, United Kingdom}
\author{I.~A.~Walmsley}
\affiliation{Clarendon Laboratory, University of Oxford, Parks Road, Oxford OX1 3PU, United Kingdom}
\author{M.~Stobi\'nska}
\email[Corresponding author, e-mail:\ ]{magdalena.stobinska@gmail.com}
\affiliation{Institute of Theoretical Physics, University of Warsaw, ul.\ Pasteura 5, 02-093 Warsaw, Poland}

\begin{abstract}	
Quantum simulations are becoming an essential tool for studying complex phenomena, e.g. quantum topology, quantum information transfer, and relativistic wave equations, beyond the limitations of analytical computations and experimental observations. To date, the primary resources used in proof-of-principle experiments are collections of qubits, coherent states or multiple single-particle Fock states. Here we show the first quantum simulation performed using genuine higher-order Fock states, with two or more indistinguishable particles occupying the same bosonic mode. This was implemented by interfering pairs of Fock states with up to five photons on an interferometer, and measuring the output states with photon-number-resolving detectors. Already this resource-efficient demonstration reveals new topological matter, simulates non-linear systems and elucidates a perfect quantum transfer mechanism which can be used to transport Majorana fermions.
\end{abstract}

\maketitle

\section{Introduction}

Quantum simulations boost the development of topological materials~\cite{Song2019}, quantum transport~\cite{Harris2017} and quantum algorithms~\cite{Jordan2012} for the benefit of low-power electronics~\cite{Khang2018}, spintronics~\cite{Smejkal2018} and quantum computing~\cite{Nayak2008}. They employ intricate quantum interference of light or matter particles. This is a challenging task: the difficulty arises from the fundamental constraint that all interfering quanta must be indistinguishable~\cite{Slussarenko2019}.  Violating this demand precludes the observation of such coherent phenomena in larger scales, in terms of particle number and duration.

So far, protocols have mainly relied on the use of three distinct quantum states: numerous qubits implemented by superconducting circuits~\cite{Houck2012} and electronic states of trapped ions~\cite{Blatt2012}; coherent states of photons~\cite{Schreiber2012} and atoms (Bose--Einstein condensates)~\cite{Fischer2004}; and multiple single-particle Fock (number) states distributed among many modes in photonic waveguides~\cite{Peruzzo2010} and optical lattices~\cite{Bloch2012}. Thus, simulations have never seriously profited from interference of multi-particle Fock states, even though the importance of this regime has been recognised~\cite{Flamini2019}, and the first attempt to mimic it with many-body systems was made~\cite{Islam2015}.

\begin{figure*}[tbp]\centering
	\includegraphics[width=12cm]{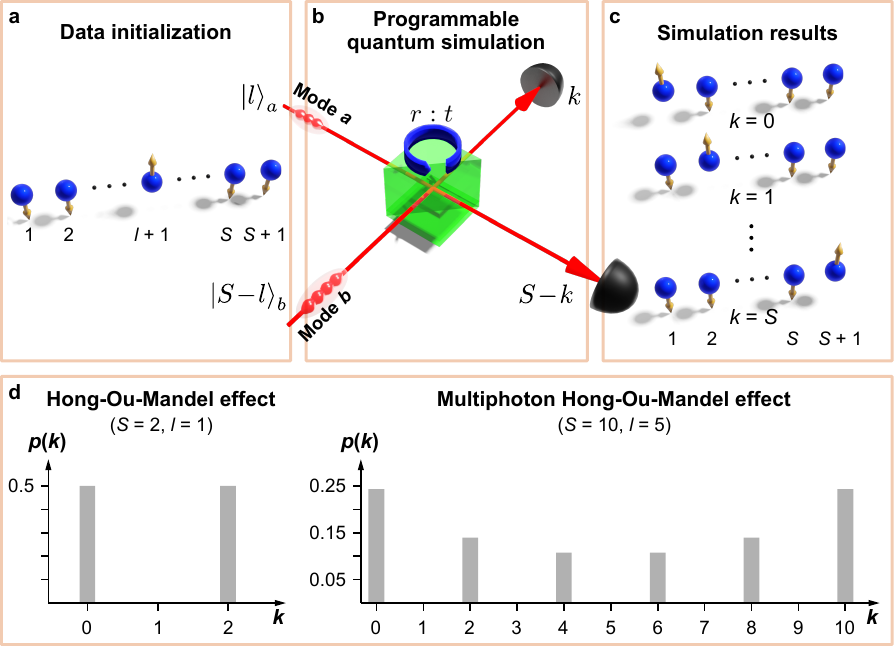}
	\caption{{\bf Photonic quantum simulations with Fock state interference.} a) Fock states $\ket{l}_a\ket{S\!-\!l}_b$ encode $(l\!+\!1)$th spin excitation in a $\frac{1}{2}$-spin chain with $S+1$ sites. b) A beam splitter with reflectivity $r$ models an XY-type of interaction in the chain that lasts $2\arcsin\sqrt{r}$. c) The likelihood of detecting $k$ and $S-k$ photons in its output ports $p(k)$, simulates the excitation probability of the $k$th spin in the chain as a result of the interaction. d) The statistics $p(k)$ originates from fundamental indistinguishability of several scenarios that occur to interfering Fock states which classically are exclusive but quantum-mechanically are coexisting, and amount to the same partitioning of incoming photons into two exit ports. Events for which quantum probability amplitudes add up non-destructively are registered more often than others.} 
	\label{fig:1}
\end{figure*}

Here we experimentally and theoretically demonstrate that multiphoton Fock state interference can be useful for quantum simulations that address applications of high impact. Remarkably, this approach grants access to a non-linearity induced by photon number detection~\cite{Scheel2003} and also avoids error accumulation that weakens methods using quantum walks, built on numerous steps~\cite{Preskill1998}. Our idea, shown in Fig.~\ref{fig:1}a-c, is based on overlapping two multiphoton Fock states, $\ket{l}_a$ and $\ket{S\!-\!l}_b $ ($l$~photons in mode~$a$ and $S-l$~in mode~$b$), on a beam splitter with tunable reflectivity $r$ which programs the simulation duration. We then collect photon statistics at its outputs. 

The primary example of a system we can simulate is a chain of $S+1$ two-level spins that initially contains just one spin excited, and that is subjected to an XY interaction. The excitation probabilities at its sites after the interaction duration are determined by the output photon statistics. These mappings are based on a solid mathematical grounding known as the Schwinger representation which links quantum harmonic oscillators with representations of spin Lie algebra $\text{su}(2)$. 

Our platform also allows us to simulate certain classes of fermionic systems, e.g.\ a non-linear Su--Schrieffer--Heeger (SSH) model~\cite{SSH1979}, obtained from the XY spin chain by a Jordan--Wigner transformation. Furthermore, we can map to Bogoliubov--de~Gennes Hamiltonians, simulating many body systems beyond the single excitation subspace e.g.\ a p-wave superconducting chain (Kitaev model)~\cite{Kitaev2001}, and the transverse-field Ising model.

Due to recent advances in photon-number-resolved detection, we were able to employ\break transition-edge sensors (TESs)~\cite{Gerrits2011} to count photons exiting the beam splitter. Amazingly, TES measurements correspond to single-site-resolved detection in the chain. The use of TESs is crucial, as Fock state quantum interference is evidenced by photon bunching. For example, two identical photons impinging on a balanced beam splitter leave in a superposition of two-photon Fock states, with both always being detected in the same output port. This is known as the Hong--Ou--Mandel (HOM) effect~\cite{Hong1987} whose generalised form can be observed for higher-order Fock states if they are prepared in similar polarisation, spectral and spatio-temporal modes~\cite{Campos1989}, as shown in Fig.~\ref{fig:1}d. 

\begin{figure*}[btp]\centering
	\includegraphics[width=12cm]{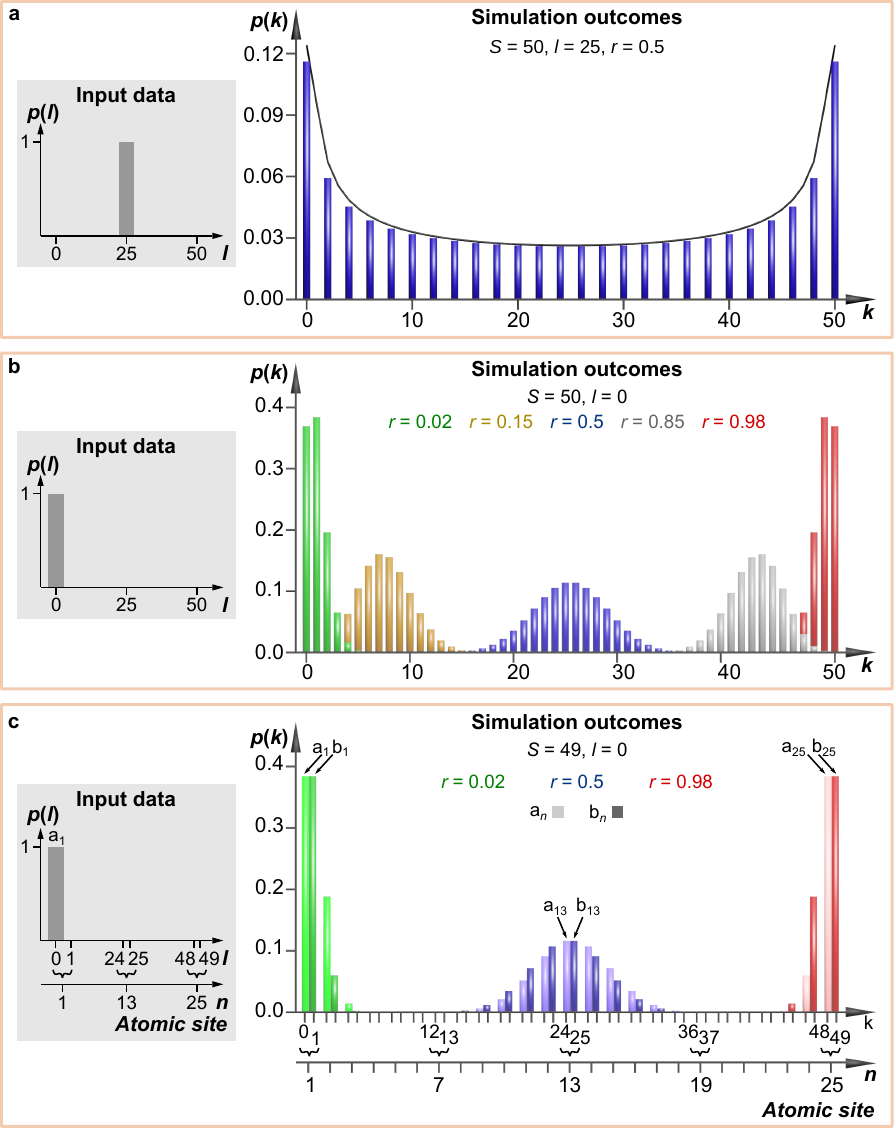}
	\caption{{\bf Encoding the outcomes of quantum simulations in photon statistics.} We repeatedly overlap two Fock states on a beam splitter of reflectivity $r$ to collect the photon statistics in its exit ports, $p(k)$. They directly provide the results of computation carried out by quantum interference. a) The first program uses $\ket{\tfrac{S}{2}}_a\ket{\tfrac{S}{2}}_b$ and $r=0.5$, revealing weakly localised edge states with a non-decaying envelope (black line), which closely resemble topological states in a non-linear SSH model. b) The second program is run for $\ket{0}_a\ket{S}_b$ and several values of $r$, demonstrating that perfect quantum wave packet transfer in a linear register results from mirror reflection of the input state w.r.t.\ the register centre. c) For an odd $S$, this program additionally simulates the perfect transfer of Majorana fermions $a_n$ and $b_n$ in a p-wave superconductor over $\tfrac{S+1}{2}$ atomic sites. The bars located at even $k$ (light green, blue, and red) correspond to the mode $a_n$ with $n=k/2 +1$, while those located at odd $k$ (dark green, blue, and red) to the mode $b_n$ with $n=(k+1)/2$.}
	\label{fig:2}
\end{figure*} 

\section{Results}

The Fock state quantum simulations build on a beam-splitter interaction $U_{\text{BS}}^{(r)}\!=\!e^{-i \theta(r) H_{\text{BS}}}$, guided by the Hamiltonian
\begin{equation}
H_{\text{BS}} = \tfrac{1}{2}(a^{\dagger}b + ab^{\dagger}),
\end{equation}
where $a^\dagger$ and $b^\dagger$ denote photonic creation operators that act on the interferometer input modes. The reflectivity $r$, defined as the probability of reflection of a single photon, encodes the interaction time $\theta(r)=2\arcsin\sqrt{r}$. For entries $\ket{l}_a$ and $\ket{S-l}_b$, the computational output from the beam-splitter and detectors is~\cite{Stobinska2019}
\begin{equation}
p(k) = \bigl\vert \braket{k, S-k | U_{\text{BS}}^{(r)} | l, S-l}\bigr\vert^2=\bigl\vert \phi_k^{(r)}(l-Sr,S) \bigr\vert^2,
\label{statistics}
\end{equation}
where $\phi_k^{(r)}(x,S)$ is the Kravchuk function~\cite{Atakishiyev1997}.

We selected three distinct examples of simulations, shown in Fig.~\ref{fig:2}, for experimental demonstration. The first \textbf{a}, uses input data initialised to $\ket{\tfrac{S}{2}}_a\ket{\tfrac{S}{2}}_b$ and the setting of $r=0.5$. For the second and third \textbf{b} \& \textbf{c}, we set $\ket{0}_a\ket{S}_b$ and repeated the computation several times whilst gradually increasing~$r$. While for the second program one can use any value of $S$, the third one runs exclusively for an odd number of photons.

{\bf Edge states in non-linear systems.} Interpretation of the outcomes of our quantum programs becomes straightforward if we consider matrix representations of $H_{\text{BS}}$ and of the Hamiltonian describing a general chiral XY $\tfrac{1}{2}$-spin chain $H_{\text{XY}} = \sum_{n=1}^{S+1} \tfrac{J_n}{2}\, (\sigma^x_n \sigma^x_{n+1} + \sigma^y_n \sigma^y_{n+1})$, where $\sigma_n^x$ and $\sigma_n^y$ are the Pauli operators acting on the $n$th spin. In the single excitation subspace spanned by the states $\ket{n} = \sigma_n^+ \ket{\downarrow_1, \dots, \downarrow_{S+1}}$, where $\sigma_n^+ = (1/2) \left( \sigma_n^x + i \sigma_n^y \right)$ is the raising operator, the latter has matrix elements $\left[ \mathbf{H}_{\text{XY}} \right]^\text{Spin}_{mn} = \braket{m | H_{\text{XY}} | n} = J_{n-1} \delta_{n,m+1} + J_{m-1} \delta_{m,n+1}$, where $\delta_{i,j}$ denotes the Kronecker delta. The elements of $H_{\text{BS}}$ in the Fock state basis are given by $\left[ \mathbf{H}_\text{BS} \right]^\text{Fock}_{nm} = \braket{n,S-n | H_\text{BS} | m,S-m}$. The two representations are identical, $\left[ \mathbf{H}_\text{BS} \right]^\text{Fock}_{nm} = \left[ \mathbf{H}_{\text{XY}} \right]^\text{Spin}_{nm}$, when we set the spin couplings to $J_n = \frac{1}{2}\sqrt{n(S+1-n)}$. As these amplitudes are non-periodic, this chain lacks translational invariance. This precludes the usual Fourier space methods used for characterising topological insulators. Remarkably, photon statistics measured behind the beam splitter is capable of simulating this non-crystalline system. The existence of topologically non-trivial states is indicated here by the fact that the Hamiltonian belongs to the chiral orthogonal (BDI) class of Altland--Zirnbauer symmetry classes, characterised by a $\mathbb{Z}$ topological invariant. Our first program performs a real-space study of this system and computes probabilities that describe its zero-energy eigenmode, $\lvert \Psi_0\rangle = \sum_{k=0}^{S} e^{-i\frac{\pi}{2}(S/2-k)}\,\phi_k^{(1/2)}(0,S)\,\sigma_{k+1}^+\ket{\downarrow_1, \dots, \downarrow_{S+1}}$. Unlike the typical edge states which are exponentially peaked at the ends of a quantum wire, these two edge states are weakly localised and plateau to a constant value in the bulk, given by $\frac{4}{\pi S\sqrt{1-(2k/S-1)^2}}$, as outlined in Fig.~\ref{fig:2}a. The intensity-dependent amplitudes $J_n$ render $H_{\text{XY}}$ a generalisation of the seminal Su--Schrieffer--Heeger (SSH) model~\cite{SSH1979} to the non-linear regime~\cite{Gorlach2017}. See Supplementary Material for details. 

{\bf Perfect state transfer.} The XY spin chain with these couplings has been extensively studied in the literature due to its remarkable property of facilitating the perfect transfer of an arbitrary quantum state~\cite{Christandl2004}. Our quantum simulation provides new insight into this system from which the perfect transfer becomes self-evident. The equivalence of $H_{\text{BS}}$ and $H_{\text{XY}}$ matrix representations implies the correspondence between interactions generated by these Hamiltonians, $U_{\text{BS}}^{(r)}$ and $U_{\text{XY}}(t)=e^{-i t H_{\text{XY}}}$, respectively. Mathematically, the beam-splitter interaction in the Fock state basis amounts to an $\alpha$-fractional Quantum Kravchuk--Fourier transform ($\alpha$-QKT) of the input state with fractionality $\alpha = \frac{4}{\pi}\arcsin\sqrt{r}$\kern.25em~\cite{Stobinska2019}. As $2$-QKT is the spatial inversion operator~\cite{Atakishiyev1997}, so is $U_{\text{XY}}(t)$ at $t=\pi$. Therefore, the transfer is an effect of mirror reflection of a quantum state w.r.t.\ the chain centre. Proving this fact was tricky within the framework of spin chains, whereas it is an evident conclusion from our photonic simulations. We note that $\alpha=2$ implies interference on a perfectly reflecting beam splitter ($r=1$) which swaps input states at its outputs. To demonstrate this behaviour, in our second program, we simulated the state transfer of a strongly localised edge state, typical of e.g.\ the SSH model. The initial Fock state $\ket{0}_a\ket{S}_b$ is gradually transformed to $\ket{S}_a\ket{0}_b$ for increasing $r$, as shown in Fig.~\ref{fig:2}b.

{\bf Generalised Majorana modes.} Multiphoton Fock state interference also facilitates the simulation of many-body systems that are not restricted to a single excitation subspace. For example, a p-wave superconducting chain (Kitaev model)~\cite{Kitaev2001} is described by the mean field Hamiltonian $H_\textrm{K} = \sum_{n=1}^{N}\{ -\mu_n (c_n^\dagger c_n - 1/2) -t_n (c_{n+1}^\dagger c_n + c_n^\dagger c_{n+1})$ $ + \Delta_n (c_{n+1}^\dagger c_n^\dagger + c_n c_{n+1}) \}$, where $c_n^\dagger$ and $c_n$ are creation and annihilation operators for electrons on the $n$th atomic site, while $\mu_n$, $t_n$ and $\Delta_n$ are site dependent chemical potentials, hopping amplitudes and superconducting pairing potentials respectively. This Hamiltonian may be expressed in the form $H_\textrm{K} = \tfrac{1}{2}\chi^\dagger \mathbf{H}_\textrm{BdG} \chi$ where $\chi = \frac{1}{\sqrt{2}}(a_1, -i b_1, a_2, -i b_2, \ldots a_{N}, -i b_{N})^\textrm{T}$ is a Nambu spinor and $\mathbf{H}_\textrm{BdG}$ is the Bogoliubov--de~Gennes Hamiltonian matrix, in the basis of Majorana operators $a_n = c_n + c_n^\dagger$ and $b_n = i(c_n^\dagger-c_n)$. The beam splitter Hamiltonian in the Fock state basis $\mathbf{H}_\text{BS}$ is identical to $\mathbf{H}_\textrm{BdG}$ for the parameters $\mu_n = J_{2n-1}, t_n = \Delta_n = \tfrac{J_{2n}}{2}$, where $2N = S+1$. This correspondence allows one to simulate the Heisenberg evolution of the Majorana operators over the interaction time $\theta(r)$, as well as the evolution of the real fermion operators $c_n$ and $c_n^\dagger$, by using linear superpositions of Fock states as input. In particular, the evolution of the operators $a_n$ and $-i b_n$ is encoded by the evolution of the photonic modes $\ket{2(n-1)}_a\ket{S-2(n-1)}_b$ and $\ket{2n-1}_a\ket{S-(2n-1)}_b$ respectively. To evidence this, a further simulation with input $\ket{0}_a\ket{S}_b$ was performed, where $S$ is an odd number, modelling the perfect transfer of Majorana modes between the two ends of a p-wave chain of $N = \tfrac{S+1}{2}$ atomic sites that is depicted in Fig.~\ref{fig:2}c. This is half the number of sites as in the XY spin chain, reflecting the fact that each physical fermion comprises a pair of Majoranas. The simulated dynamics also apply to one-dimensional arrays of photonic cavities~\cite{Bardyn2012} where the effective superconducting pairing and Majorana modes arise from Kerr-type non-linearities within a Bose--Hubbard model.

{\bf Non-uniform transverse-field Ising chain.} One can also simulate a transverse-field Ising model, $H_\textrm{K} = \frac{1}{2}\sum_{n=1}^N(\mu_n \sigma_n^z + 2t_n \sigma_n^x \sigma_{n+1}^x)$, since this is related to the p-wave superconducting chain by a Jordan--Wigner transformation. Due to the non-uniform field $\mu_n$ and spin couplings $t_n$, the system inherits the perfect mirror reflection from the beam splitter dynamics and allows for perfect state transfer after an interaction time $\theta = \pi$. We thus highlight a new quantum spin network that allows perfect transfer, similar to the previously discussed XY model, but which has not been considered by previous authors. For an example, to simulate the transfer of an excited spin between ends of a chain, one should interfere the state $\frac{1}{\sqrt{2}}(\ket{0}_a\ket{S}_b + \ket{1}_a\ket{S-1}_b)$ on a balanced beam splitter. See Supplementary Material for details.

\section{Experimental study}

\begin{figure*}[tbp]\centering
	\includegraphics[width=12cm]{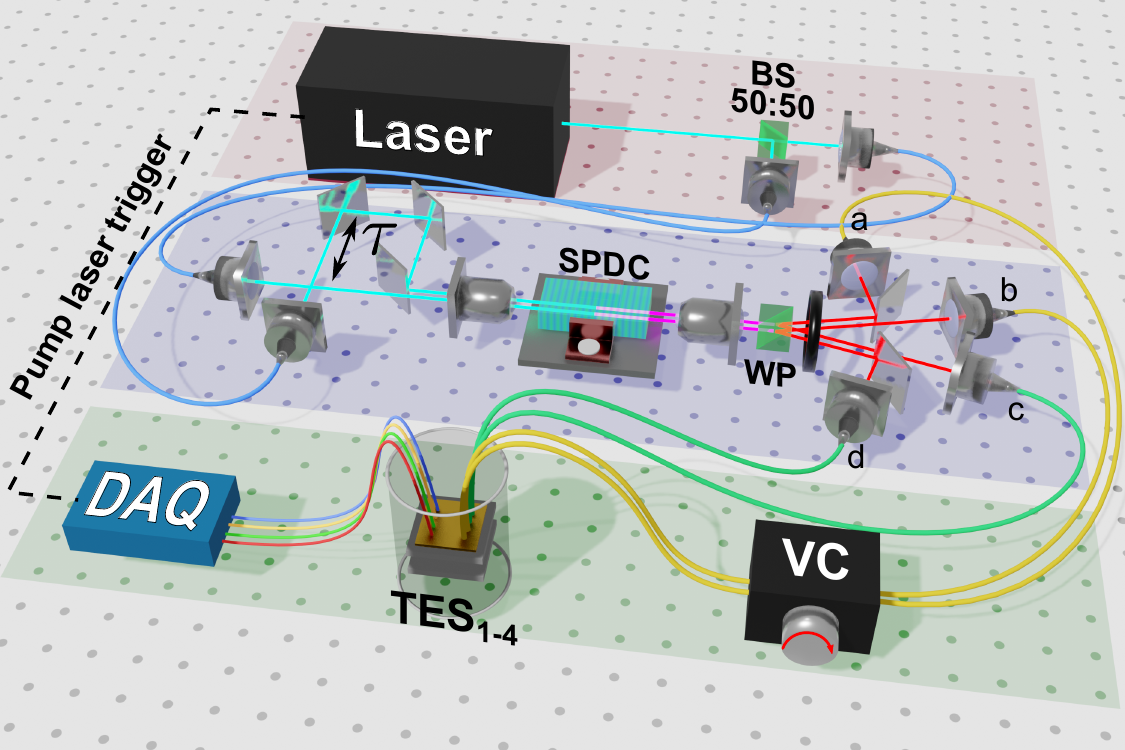}
	\caption{{\bf Experimental integrated-photonics set-up for Fock state interference.} Laser pulses (blue beams) centred at $775\kern.25em\mathrm{nm}$ pump two collinear type-II phase-matched $8\kern.25em\mathrm{mm}$-long spontaneous parametric down-conversion (SPDC) waveguides written in a periodically poled KTP (PP-KTP) crystal. Each SPDC creates a two-mode photon-number correlated state (red beams). The modes are separated with a Wollaston prism (WP) into the modes $a$--$d$. They are filtered by bandpass filters tuned to the central wavelength $1554\kern.25em\mathrm{nm}$ for the signal modes $a$ and $b$, and $1546\kern.25em\mathrm{nm}$ for the idler modes $c$ and $d$. The idler beams are used for heralding the creation of the signal Fock states in $a$ and $b$ which interfere in a variable ratio phase-matched fibre coupler (VC). The VC allows us to set the ratio between $0$ and $1$ with an error of $\pm1.5\times 10^{-2}$. We used transition-edge sensors (TESs) with efficiency exceeding $90\%$ for photon-number-resolved measurements in all modes~\cite{Gerrits2011}. The optimal temporal overlap at the VC is achieved by adjusting an optical path delay~$\tau$. The data is analysed with a data acquisition unit (DAQ).}
	\label{fig:3}
\end{figure*}

Fig.~\ref{fig:3} shows the experimental integrated-photonics schema used for Fock state quantum simulations. Two pulsed spontaneous parametric down-conversion sources (SPDC) each generated independent two-mode photon-number-entangled states $\ket{\Psi} = \sum_{n=0}^{\infty} \sqrt{\braket{n}^n}\,\ket{n,n}$ with an average photon number $2\braket{n}=0.4$. For the pump repetition rate of $75\kern.25em\mathrm{kHz}$ this led to approximately $0.46$ five-photon ($12$ four-photon) Fock states created per minute in each arm of the SPDC, of which about 0.2 (6) reached the detectors due to ca. $50\%$ losses in the set-up. One mode from each $\ket{\Psi}$ (the idlers, $c$ and $d$) was sent to a TES. Due to photon-number entanglement in $\ket{\Psi}$ states, the outcomes of TESs, $l$ and $S-l$, heralded the creation of Fock states $\ket{l}_a$ and $\ket{S-l}_b$ in the signal modes $a$ and $b$. 

We characterised the set-up to confirm the high degree of indistinguishability of these Fock states, the key issue for multiphoton HOM effect. We measured the standard HOM interference dip between both sources for a small mean photon number of the order of $10^{-4}$, and achieved the visibility $V_\textrm{HOM}=85.9\%$. Next, we took a measurement of the second order correlation function for each SPDC source separately and observed $g^{(2)}\geq1.86\approx1+V_\textrm{HOM}$, which corroborates the previous result. An effective Schmidt mode number of $K=\frac{1}{g^{(2)}-1}=1.16$ proves our SPDC sources nearly single-mode.

\begin{figure*}[tbp]\centering
	\includegraphics[width=12cm]{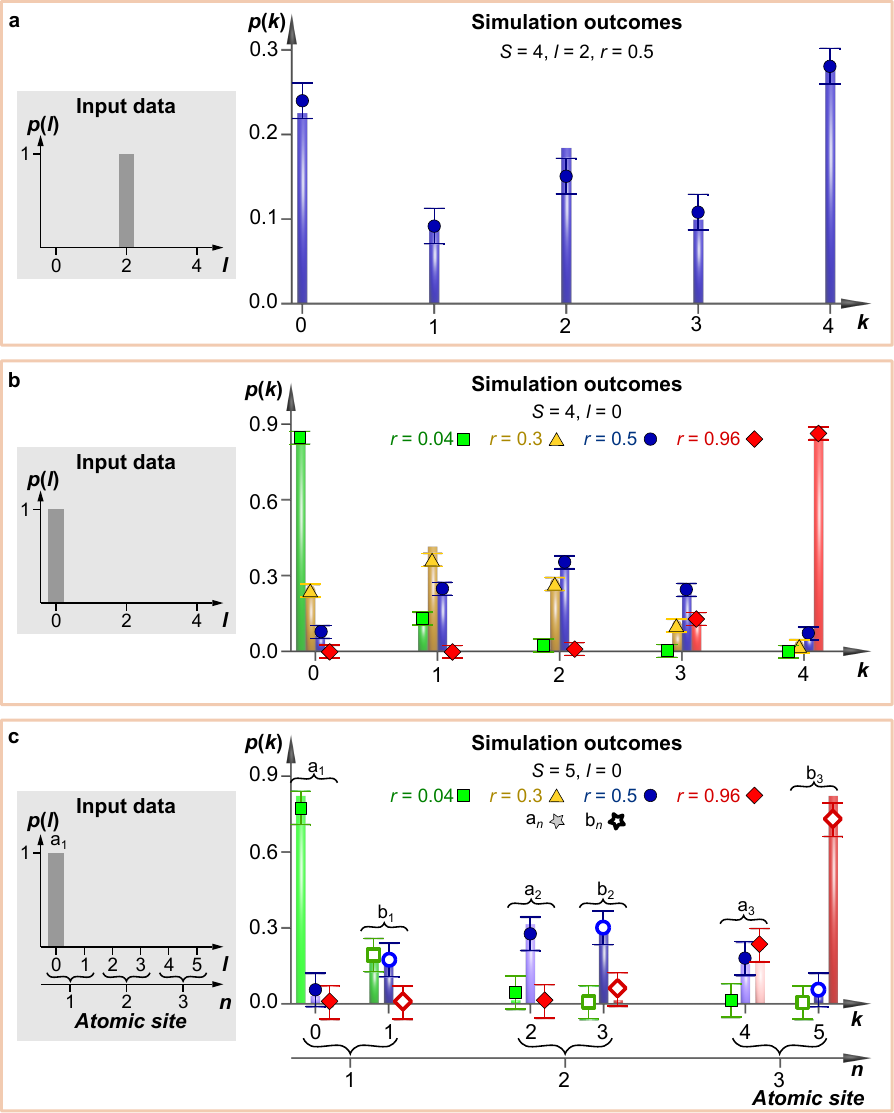}
	\caption{{\bf Measured Fock state quantum simulations.} The experimental simulation outcomes (symbols with error bars) are the directly measured photon statistics $p(k)$ resulting from interference of Fock states, $\ket{l}_a$ and $\ket{S-l}_b$, on a beam splitter of reflectivity $r$. The grey insets display the input data $l$, while the bars expected theoretical values of $p(k)$ obtained numerically. a) Two two-photon Fock states ($S=4, l=2$) interfering on a balanced beam splitter ($r=0.5$) reveal an edge-state structure in the corresponding spin chain. b) Interference of the vacuum and a four-photon number state ($S=4, l=0$) observed for several gradually increasing reflectivities $r=0.04$ (green squares), $0.3$ (orange triangles), $0.5$ (blue circles), and $0.96$ (red diamonds) models perfect wave packet transfer in the Kravchuk chain. c) Interference of the vacuum and a five-photon number state ($S=5, l=0$) observed for the same set of reflectivities models the transfer of Majorana fermions $a_n$ and $b_n$ across a p-wave superconducting chain of $3$ sites.}
	\label{fig:4}
\end{figure*}

The measured simulations are presented in Fig.~\ref{fig:4}. The data shown in Fig.~\ref{fig:4}a and Fig.~\ref{fig:4}b consists of approximately $1.6 \times 10^3$ registered events, for each value of $r$, in which the total number of photons was $S=4$. The data in Fig.~\ref{fig:4}c comprises $2.3\times 10^2$ measurements, for each value of $r$, in which $S=5$. We compared them with a numerical model based on Eq.~(\ref{statistics}) supplemented with the analysis of experimental imperfections, and found that they are in a good agreement. Errors were estimated as a square root inverse of the number of measurements. See Methods for details.

In Fig.~\ref{fig:4}a we show the photon statistics recorded by TES$_{2-3}$ for the coupler splitting ratio $r=0.5$, conditioned on the heralded photon numbers $l=2$ and $S-l=2$ in modes $c$ and $d$. They directly model a zero-energy eigenmode of a non-linear SSH model described by $\left[ \mathbf{H}_\text{BS} \right]^\text{Fock}_{nm}$, with emerging two weakly localised edge states. Fig.~\ref{fig:4}b depicts the statistics gathered for $l=0$ and $S-l=4$ for several splitting ratios: $r=0.04$ (green squares), $0.3$ (orange triangles), $0.5$ (blue circles) and $0.96$ (red diamonds). It visualises perfect state transfer of the first spin excitation in the chain of $5$ particles by means of continuous-time mirror reflection w.r.t.\ the chain centre. Fig.~\ref{fig:4}c shows an experimental simulation of the perfect transfer of a Majorana fermion in a p-wave superconducting chain of 3 sites that is based on the statistics gathered for $l=0$ and $S-l=5$ for all the listed values of $r$.

\section{Conclusions and outlook}

Multi-particle Fock state interference is a new and compelling method in the field of quantum simulations, promising for studying non-crystalline topological materials, beyond the recently challenged bulk-edge correspondence theorem~\cite{Downing2019}. It allowed us to simulate systems as diverse as an XY spin chain and a non-linear SSH model, as well as the perfect transfer of Majorana fermions over a quantum wire, in a system that is not tied to a single-excitation subspace. Importantly, photon-number-resolved detection introduces an effective non-linearity~\cite{Scheel2003} which can be harnessed in simulated models. The presented examples apply to a variety of systems such as superconducting nanowires~\cite{Diez2012}, disordered graphene quasi-1D nanoribbons~\cite{Han2007} and disordered cold atoms~\cite{Pinheiro2015}. These may find applications in next-generation electronics~\cite{McCaughan2014} and spintronics~\cite{Nautiyal2004} operating with almost no energy dissipation and speeds exceeding 100\kern.25emGHz. Our simulations amount to computations of the Kravchuk transform that classically is expensive but in the quantum domain can be attained with a single gate~\cite{Stobinska2019}.

Multiphoton Fock states have been utilised in quantum simulations in a very limited capacity until now. In photonics, the main focus was on successful manipulation of large numbers of single or pairs of photons in bulk optics~\cite{Zhong2018}, as well as in integrated platforms~\cite{Wang2019}. For example, only one- and two-photon output states of a quantum walk in coupled waveguides were measured, which are a small fraction of the total output~\cite{Peruzzo2010}. The advantage of these photonic systems, however, lies in easily engineered waveguide layouts which can be used to e.g.\ model different couplings in the chains. Nevertheless, keeping a high degree of indistinguishability of photons coming from different sources remains a challenge~\cite{Slussarenko2019}.

On the contrary, our simulations are the first to be done exclusively in Fock space, with Fock states of high photon number encoding all the information from input to output. Although currently the experimental generation of five-photon Fock states is already beyond the state of the art, it is soon expected to reach the level of tens of photons~\cite{Harder2016}. Moreover, our method avoids some of the error accumulation and scaling problems of the waveguide-based set-ups. It can also be extended to higher dimensions by including additional degrees of freedom such as photon frequency and polarisation. The scope of simulations could be further broaden by using input states superpositions $\sum_{l=0}^S x_l \ket{l,S-l}$ and altering the spin-chain couplings. Although preparation of such general superpositions poses a challenge in photonics, input states in the form of generalised Holland--Burnett states were experimentally obtained by interfering Fock states on a beam splitter~\cite{Thekkadath2020}. Some other examples could be reached by heralding and conditional state preparation using more intricate interferometers. Merging our approach with coupled-waveguide set-ups is yet an unexplored and intriguing territory.

It would also be very interesting to implement our technique with quantum simulation platforms that are universal. For example, Fock states are also available in motional states of trapped ions up to 10 excitations~\cite{Ding2017} and in the form of plaquette Fock states of atoms in optical lattices up to 4 excitations~\cite{Islam2015}. The range of accessible parameters controlling these systems could provide access to other complementary simulation models. Moreover, deterministic creation of an arbitrary superposition of Fock states has been demonstrated for trapped ions and superconducting resonators~\cite{Hofheinz2009}. This would further expand the assortment of input states that could be used for simulation and may give birth to new fascinating results.

\section{Methods}

\subsection{Characterisation of the set-up}

Each integrated SPDC source produced a two-mode weakly squeezed vacuum state $\ket{\Psi} = \sum_{n=0}^{\infty} \lambda_n\,\ket{n,n}_{s,i}$, where $s$ and $i$ denote two output modes, named the signal and idler, $\lambda_n = \tfrac{\tanh^n g}{\cosh g}$, $\lvert \lambda_n\rvert^2$ is a probability of creation of a pair of $n$ photons and $g$ is the parametric gain. The average photon number in each mode of $\ket{\Psi}$ is $\braket{n} = \sinh^2 g$. The observed average photon number of $\braket{n} \approx 0.2$ amounts to $g=0.44$, which was sufficient to ensure the emission of multiphoton pairs. In this regime one can approximate $\cosh g \approx 1$ and thus, $\lambda_n \approx \sinh^n g = \sqrt{\braket{n}^n}$.

The transition-edge sensors (TESs) were operated at $70\kern.25em\mathrm{mK}$, which allowed photon-number resolved measurements in all modes~\cite{Gerrits2011}.

The transmission losses in the set-up were estimated by means of Klyshko efficiency measurements. To this end, we set the reflectivity of variable coupler at $r=0.5$, and pumped each of the two SPDC sources separately at successively lower power values. The registered four-mode photon statistics were then binned into `photon(s)/no-photon' datasets to mimic the use of standard binary detectors, e.g.\ avalanche photo-diodes, and we concluded the total efficiencies of the heralding modes $c$ and $d$ to be $\eta_c=50.3\%$ and $\eta_d=48.5\%$, respectively. The variable-coupler modes $a$ and $b$ exhibited a total efficiency of $\eta_a=21.6\%$ and $\eta_b=20.6\%$, respectively. These values result from the fact that each mode carried a $3\kern.25em\mathrm{dB}$ loss from the coupler itself and another $1\kern.25em\mathrm{dB}$ due to coupler insertion and fibre-to-fibre coupling losses. We estimated the transmission losses to be approximately of $50\% \approx 3\kern.25em\mathrm{dB}$. Here $1\kern.25em\mathrm{dB}$ stands for the initial fibre in-coupling loss due to spatial mode mismatch, while $0.25\kern.25em\mathrm{dB}$ stems from detectors inefficiencies, and the remaining loss is from three FC/PC fibre-to-fibre couplers per mode as well as bending losses in the transmission fibres between the set-up and the detectors.

The HOM visibility is computed using the formula
$v^{(2)} = \frac{n_{\text{max}} - n_{\text{min}}}{n_{\text{max}} + n_{\text{min}}}$, where $n_{\text{max}}$ and $n_{\text{min}}$ are the maximal and minimal number of events registered by the TES detectors for the given photon number $S$. In the experiment for input $\ket{2,2}$ and $r=0.5$, we obtained $v^{(2)}= 50.6\%\pm1.2\%$, whereas for input $\ket{0,4}$ ($\ket{0,5}$) they were $99.1\%\pm2.5\%$ ($97.8\%\pm6.2\%$) for $r=0.04$, $87.6\%\pm2.2\%$ ($96.7\%\pm7.2\%$) for $r=0.3$, $65.7\%\pm1.7\%$ ($71.4\%\pm4.6\%$) for $r=0.5$ and $99.9\%\pm0.8\%$ ($98.6\%\pm7.2\%$) for $r=0.96$. 

\subsection{Error estimation}

In the experiment, each measurement results in a 4-tuple consisting of the number of photons registered by TES${}_{1-4}$, corresponding to photon-number states in modes $a$-$d$ (Fig.~\ref{fig:3}). The tuple counts are stored in a database. The probability of detecting $k$ and $S-k$ photons in modes $a$ and $b$ is computed as $p(k)=N_k/N$, where $N_k$ is the database value retrieved for the key $(k,S-k,l,S-l)$ and $N$ is the total count of events characterised by the given total number of photons $S$. The measurement errors for each mode were estimated to $\Delta p=1/\sqrt{N}$.

\subsection{Numerical model of experimental outcomes}

To assess the experimental results we developed a theoretical model which extended Eq.~(\ref{statistics}) by taking into account the influence of losses, multi-modeness of beams as well as inefficient photodetection. 

Decoherence resulting from the first two effects was modelled by replacing the mode $b^{\dagger}$ with a superposition of the same mode  $b^{\dagger}$ and an orthogonal one $b^{\dagger}_{\perp}$, i.e.\ $b^{\dagger} \to \cos y \, b^{\dagger} + \sin y \, b^{\dagger}_{\perp}$, where the parameter $y \in (0,\tfrac{\pi}{2})$ introduced weights and `tuned' the distinguishability. This transformation led to the interference of $\ket{l}_a$ with a two-mode Fock state superposition\break $\sum_{n=0}^{S-l} \binom{S-l}{n}^{-1/2} \cos^n y \, (\sin y)^{S-l-n} \ket{n}_b \ket{S-l-n}_{b\perp}$ instead of the single-mode Fock state\break $\ket{S-l}_b$, as before. Thus, effectively, some of the multiphoton states interfered with the vacuum state and this implemented the usual model describing particle loss. In our computations, we took $y = \arcsin\sqrt{(K-1)/K}$, where $K$ denoted the effective Schmidt mode number measured during the set-up characterisation. For $K=1.16$, we used $y=0.38$. 

Realistic model of photodetection requires taking into account a probability of detecting $n_d$ photons when a Fock state $\ket{n_{\text{in}}}$ reaches a TES. It is given by $p_{\text{TES}}(n_{\text{in}}, n_d) = \binom{n_{\text{in}}}{n_d} (1-\eta)^{n_{\text{in}}-n_d}\,\eta^{n_d}$ where $n_d\leq n_{\text{in}}$ and $\eta$ is the detector efficiency. In our computations we first used a starting value of $\eta=0.7$ and then numerically optimised efficiencies for individual TESs to compensate for the uneven photon number distribution $p(k)$ seen in Fig.~\ref{fig:3}a. The programs were written in Python using mpmath library. 

\acknowledgments

\begin{description}

\item[Funding:]

TS, TM, AB and MS were supported by the Foundation for Polish Science `First Team' project No.\ POIR.04.04.00-00-220E/16-00 (originally: FIRST\ TEAM/2016-2/17) and the National Science Centre `Sonata Bis' project No. 2019/34/E/ST2/00273. AE and IW  were supported by the Engineering and Physical Sciences Research Council project No. EP/K034480/1.

\item[Authors contributions:]

TS, TM, AB and MS developed the theory while AE, WRC, WSK, and IAW were responsible for realisation of the experiment. JJR, SWN, TG, and AL delivered and maintained the transition edge sensor detection system. AB, TS and TM developed the software and performed numerical computations. AB prepared the plots. All the co-authors wrote up the manuscript.

\item[Competing interests:]

The authors declare that they have no competing financial interests.

\item[Data and materials availability:]

All data needed to evaluate the conclusions in the paper are present in the paper and/or the Supplementary Material. Additional data available from authors upon request.

\end{description}

\appendix\onecolumngrid

\section{The Schwinger representation: mapping between photonic and spin platforms}

It is a well-known fact in the representation theory of groups and algebras that the Lie algebra $\text{su}(2)$ can be represented in terms of annihilation and creation operators of a harmonic oscillator. This is known as the Schwinger representation~\cite{Chruscinski2004}. It allows one to associate two independent quantum-harmonic oscillator modes with spin operators as follows
\begin{equation}
	S_x=\dfrac{a^\dagger b+a\,b^\dagger}{2},\qquad
	S_y=\dfrac{i\left(a\,b^\dagger-a^\dagger b\right)}{2},\qquad
	S_z=\dfrac{a^\dagger a-b^\dagger b}{2},\qquad
	S_0=\dfrac{a^\dagger a+b^\dagger b}{2},
	\label{Schwinger}
\end{equation}
where $S_0$ is the Casimir operator $S_0 (S_0+1) = S_x^2+S_y^2+S_z^2$ and the standard $\text{su}(2)$ commutation relations hold
\begin{equation}
	[S_x, S_y] = iS_z, \qquad [S_y,S_z] = iS_x, \qquad [S_z,S_x] = iS_y.
\end{equation}
Therefore, we can immediately identify that 
\begin{equation}
	H_{\text{BS}}= S_x.
\end{equation}

In this picture, a two-mode Fock state $\ket{l}_a \ket{S-l}_b$ corresponds to spin-$\tfrac{S}{2}$ particle that is prepared in an eigenstate of $S_z$ with eigenvalue $l-\tfrac{S}{2}$, known as a Dicke state $\ket{\tfrac{S}{2}; l-\tfrac{S}{2}}$
\begin{equation}
	S_z \ket{l}_a \ket{S-l}_b = (l-\tfrac{S}{2})\ket{l}_a \ket{S-l}_b,
\end{equation}
\begin{equation}
	\ket{l}_a \ket{S-l}_b \equiv \ket{\tfrac{S}{2}; l-\tfrac{S}{2}}.
\end{equation}
Furthermore, one can one-to-one map the Dicke states $\ket{\tfrac{S}{2}; m}$ to the basis states that span the single excitation subspace of a spin-$\tfrac{1}{2}$ chain. To this end, we employ the following relabelling $m=-\tfrac{S}{2}+n-1$, where $1\le n \le S+1$ denotes $n$th spin-$\tfrac{1}{2}$ in the chain with $S+1$ sites. Then, $\ket{\tfrac{S}{2}; l-\tfrac{S}{2}}$ corresponds to the chain where $(l\!+\!1)$th spin-$\tfrac{1}{2}$ is excited
\begin{equation}
	\boxed{
		\ket{l}_a \ket{S-l}_b  \equiv\ket{\tfrac{S}{2}; l-\tfrac{S}{2}}  \equiv \ket{\downarrow_1, \dots, \uparrow_{l+1}, \dots, \downarrow_{S+1}} = \sigma_{l+1}^+ \ket{\downarrow_1, \dots, \downarrow_{S+1}},}
\end{equation}
where $\sigma_m^+ = \frac{1}{2}\left(\sigma_m^x + i \sigma_m^y \right)$ is the raising operator acting on $m$th spin, with $\sigma_m^x$ and $\sigma_m^y$ denoting the Pauli operators. For a concise notation, we denote such spin-chain states as follows 
\begin{equation}
	\ket{m} = \sigma_{m}^+ \ket{\downarrow_1, \dots, \downarrow_{S+1}}.
	\label{spin-states}
\end{equation}

\section{Introduction to Fock-state photonic quantum simulations}

The key observation that provides the basis for quantum simulations based on Fock-state interference is the formal mathematical mapping between the Hamiltonian matrix representations of an XY-type of interacting spin chain and that of a beam splitter.

As we pointed out in the main text, a general chiral XY spin chain is represented by the Hamiltonian
\begin{equation}
	H_{\text{XY}} ={} \sum_{n=1}^{S} \frac{J_n}{2}\, (\sigma^x_n \sigma^x_{n+1} + \sigma^y_n \sigma^y_{n+1}) = \sum_{n=1}^{S} J_n \left( \sigma_n^+ \sigma_{n+1}^- + \sigma_n^- \sigma_{n+1}^+ \right).
	\label{H_XY}
\end{equation}

In the single excitation subspace that is spanned by the states shown in Eq.~(\ref{spin-states}), this Hamiltonian has the matrix representation
\begin{align}
	\left[ \mathbf{H}_{\text{XY}} \right]^\text{Spin} ={}&\begin{pmatrix}
		0 & J_1& 0 &... & 0\\
		J_1 & 0 & J_2 &... & 0\\
		0 & J_2 & 0 &... & 0\\
		. & . & . & ... & . \\
		. & . & . & ... & . \\
		. & . & . & ... & J_S \\
		0 & 0 & 0 & J_S & 0\\
	\end{pmatrix},
	\label{hmatrix}
\end{align}
where the matrix elements equal $\left[ \mathbf{H}_{\text{XY}} \right]^\text{Spin}_{mn} = \bra{m} H_{\text{XY}} \ket{n}=J_{n-1}\, \delta_{n,m+1} + J_{m-1}\, \delta_{m,n+1}$ and $\delta_{i,j}$ is the Kronecker delta; $\delta_{i,j} = 1$ if $i=j$ and $\delta_{i,j} = 0$ otherwise.

The beam splitter Hamiltonian~\cite{Kim2002}
\begin{equation}
	H_{\text{BS}} = \tfrac{1}{2}(a^{\dagger}b + ab^{\dagger}),
\end{equation}
where $a^\dagger$ ($a$) and $b^\dagger$ ($b$) denote photonic creation (annihilation) operators which act on the interferometer input modes, features the following matrix representation elements in the Fock state basis
$\left[ \mathbf{H}_\text{BS} \right]^\text{Fock}_{nm} = \bra{n,S-n} H_\text{BS} \ket{m,S-m}= \frac{1}{2}\sqrt{n(S+1-n)}\, \delta_{n,m+1} + \frac{1}{2}\sqrt{m(S+1-m)}\, \delta_{m,n+1}$. Please note that in the notation used the photonic state $\ket{m,S-m}$ corresponds to spin chain state $\ket{m+1}$. Therefore, if we set the spin-couplings to 
\begin{equation}
	J_n = \tfrac{1}{2}\sqrt{n(S+1-n)},
	\label{spin_couplings}
\end{equation} 
then
\begin{equation}
	\boxed{
		\left[ \mathbf{H}_\text{BS} \right]^\text{Fock}_{nm} = \left[ \mathbf{H}_{\text{XY}} \right]^\text{Spin}_{nm}.}
\end{equation}

\section{Quantum program 1: simulation of weakly localised edge states}

\noindent
{\bf Input data initialised to:} 
$\ket{l=\tfrac{S}{2}}_a\ket{S-l=\tfrac{S}{2}}_b$.
\smallskip

\noindent
{\bf Program setting:} $r=0.5$.

\subsection{Short introduction to edge states: the Su--Schrieffer--Heeger (SSH) model}

\begin{figure}[t]\centering
	\includegraphics[width=10cm]{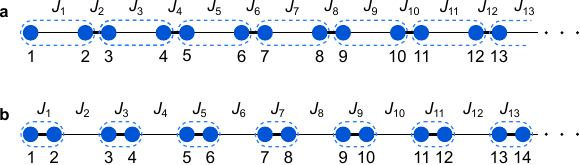}
	\caption{\textbf{The SSH model:} an infinite dimerised chain of atoms with couplings $J_n= 2J(1+\delta(-1)^n)$. It reveals translational invariance and chiral symmetry. \textbf{a}, When couplings between consecutive cells (marked with a dashed line) are stronger than intra-cell ones ($0<\delta\leq 1$), localised edge states are observed, \textbf{b},~Strong intra-cell couplings within each cell prohibit creation of such edge states ($-1\leq\delta<0$).}
	\label{fig:dimerization}
\end{figure}

\begin{figure}[h]
	\raisebox{3.5cm}{\textbf{a}}\quad
	\includegraphics[height=4cm]{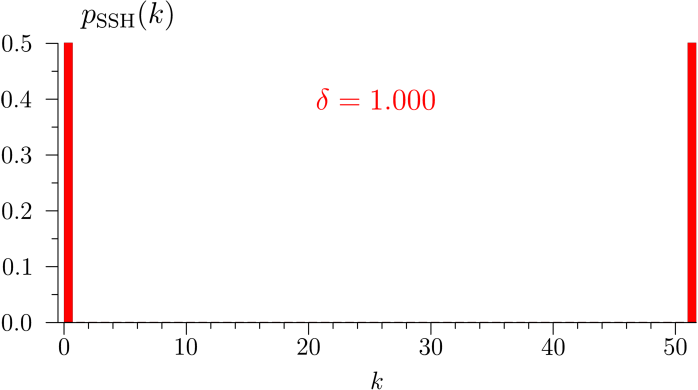}
	\hskip1cm
	\raisebox{3.5cm}{\textbf{b}}\quad
	\includegraphics[height=4cm]{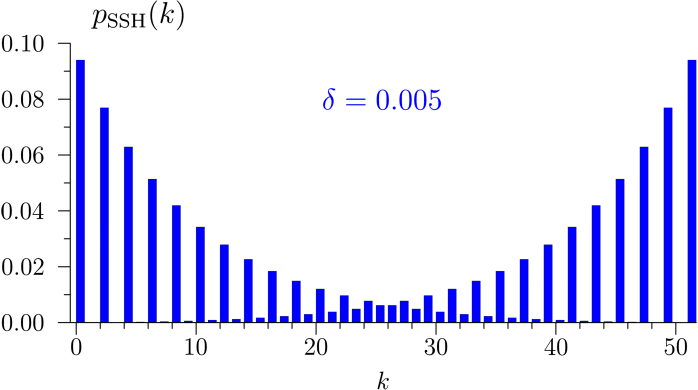}
	\caption{\textbf{Edge states in the SSH model:} probability densities of the zero-energy edge states for a chain of length $S+1=52$, computed \textbf{a}, for $\delta=1$ (red) and \textbf{b}, $\delta=0.005$ (blue).}
	\label{fig:SSHvsGeneral}
\end{figure}

The SSH model is the seminal example of a 1D system where edge (topologically non-trivial) states can be observed. The system is a dimerised chain of atoms that is described by a Hamiltonian of the form shown in Eq.~(\ref{H_XY}) with periodic couplings of $J_n = 2J(1+\delta(-1)^n)$, where $-1\le \delta \le 1$ is the dimerisation parameter. Therefore the chain consists of alternating couplings, one weaker and one stronger. The dimerisation can be chosen in such a way that the atoms at the ends of the chain experience either the weaker coupling ($\delta > 0$) or the stronger coupling ($\delta < 0$), shown in Fig.~\ref{fig:dimerization}. The first option results in the formation of topologically non-trivial states and the second one trivial states. The topological phase transition takes place at $\delta=0$.

In the topologically non-trivial phase, the system is a topological insulator with two zero-energy edge states. For example, the left boundary state is of the form
\begin{equation}
	\sum_{n=1}^{(S+1)/2} (-1)^n e^{-2n/\xi} \sigma_{2n}^+\ket{ \downarrow_1, \dots \downarrow_{S+1}},
\end{equation}
where $\xi = 2/\ln\bigl(\frac{1+\delta}{1-\delta}\bigr)$ is the localisation length~\cite{Nevado}. Usually, the boundary states are studied for $\delta$ close to $1$, when they are exponentially peaked at the ends of the chain, shown red in Fig.~\ref{fig:SSHvsGeneral}a. Fig.~\ref{fig:SSHvsGeneral}b also shows the zero-energy states but in a less studied regime, close to the topological phase transition, for $\delta=0.005$ (blue). Interestingly, the conventional SSH model can present weakly localised edge states that are nonetheless topological.

\subsection{Edge states in a non-linear SSH model}
\label{ssec:nonlinear_SSH}

\begin{figure}[t]\centering
	\includegraphics[width=10cm]{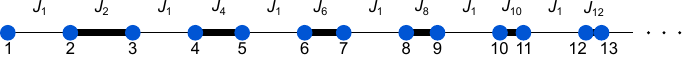}
	\caption{\textbf{A non-linear SSH model:} even couplings $J_{2n}$ are intensity-dependent, while $J_1 = \textrm{const}$. }
	\label{fig:single_edge_SSH}
\end{figure}

\begin{figure}[h]\centering
	\includegraphics[width=7cm]{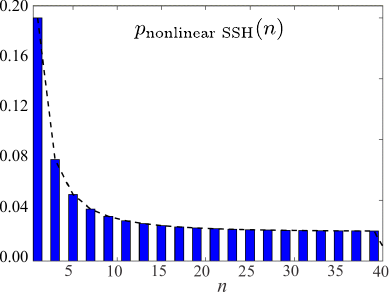}
	\caption{\textbf{An edge state in a non-linear SSH model with intensity-dependent couplings~\cite{Gorlach2017}.}
		The state reveals a non-decaying envelope. The probability distribution is calculated for $J_1 = 2.3\times 10^{-3},$ $J_2 = 2.0\times 10^{-3}$ and $\alpha = 5.0\times 10^{-5}$.}
	\label{nonlinear_SSH}
\end{figure}

An interesting generalisation of the SSH model to the non-linear domain was achieved by setting intensity-dependent site couplings~\cite{Gorlach2017}. This model was theoretically implemented with an array of cavities with the tunneling constants equal $J_{2n} = J_2 + \alpha \left(\lvert E_{2n} \rvert^2 + \lvert E_{2n+1} \rvert^2\right)$ and $J_{2n+1} = J_1= \textrm{const}$, as shown in Fig.~\ref{fig:single_edge_SSH}. Here $E_n$ denotes the field amplitude in $n$th resonator. Fig.~\ref{nonlinear_SSH} depicts a self-induced topological edge state that arises in this system. It is plotted for $J_1>J_2$, a regime where the linear SSH model shows no boundary states. Interestingly, its envelope reveals no exponential decay.

\subsection{Our simulation: edge states in a generalised non-linear SSH model}

\begin{figure}[t]\centering
	\includegraphics[width=10cm]{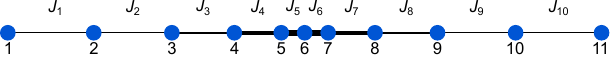}
	\caption{\textbf{Quantum simulation of a generalised non-linear SSH model:} Fock-state interference on a beam splitter mimics a finite spin chain with spatial-inversion symmetry, that belongs to the BDI class of topological insulators. The couplings $J_n=\frac{1}{2}\sqrt{n(S+1-n)}$ lead to the formation of weakly localised edge states. The figure shows a chain of $S+1=11$ spins.}
	\label{fig:generalisedSSH}
\end{figure}

Interference of Fock states on a beam splitter can simulate an interacting spin chain with non-periodic next-neighbour spin couplings shown in Eq.~(\ref{spin_couplings}) and Hamiltonian given by Eq.~(\ref{H_XY}). This system is depicted in Fig.~\ref{fig:generalisedSSH}. Interestingly, the couplings~(\ref{spin_couplings}) also are intensity dependent, as $n$ is the intensity of Fock state $\ket{n}$ and thus, we also work with a non-linear SSH-type of model. However, unlike in Section \ref{ssec:nonlinear_SSH}, the dependency is not linear.

\subsubsection*{The BDI Altland-Zirnbauer symmetry class}

In the periodic table of topological insulators defined by the Altland--Zirnbauer symmetry classes~\cite{Kitaev2009}, a system is categorised according to the properties of its time-reversal operator $T = U_T \otimes \mathcal{K}$, charge-conjugation operator $C = U_C \otimes \mathcal{K}$, and chiral-symmetry operator $\Gamma = T \otimes C$, where $U_{T(C)}$ is a unitary operator and $\mathcal{K}$ is complex conjugation. If there exists a $T$ ($C$ or $\Gamma$) that commutes (anti-commutes) with the system Hamiltonian, than the system is said to possess the respective symmetry and is classified according to the square of that operator. The beam-splitter Hamiltonian matrix representation $\left[ \mathbf{H}_\text{BS} \right]^\text{Fock}$ is real-valued and thus, its time-reversal symmetry operator is simply $T = 1 \otimes \mathcal{K}$. Due to the absence of couplings beyond nearest-neighbour the chiral symmetry operator is given by $\Gamma_{ij} = (-\delta_{ij})^2$, and finally $C = T \otimes \Gamma$. Our photonic system possesses all three symmetries with $(T^2,C^2,\Gamma^2) = (1,1,1)$. Thus it belongs to the BDI (chiral orthogonal) class of topological insulators which in one-dimension is characterised by a $\mathbb{Z}$ topological invariant. Therefore, the simulated spin chain does so too.

\subsubsection*{Weakly localised zero-energy edge states}

\begin{figure}[h]\centering
	\raisebox{3.5cm}{\textbf{a}}
	\includegraphics[height=4cm]{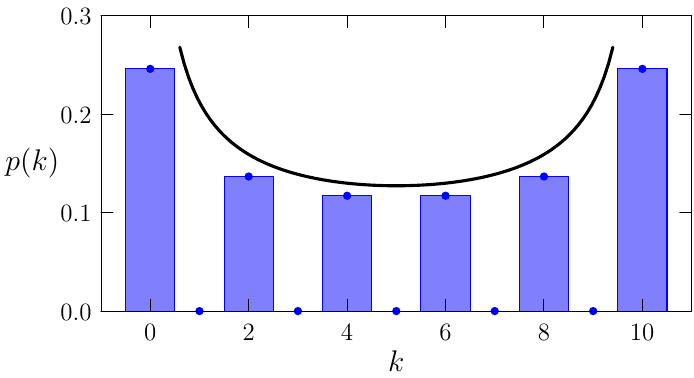}
	\hskip1cm
	\raisebox{3.5cm}{\textbf{b}}
	\includegraphics[height=4cm]{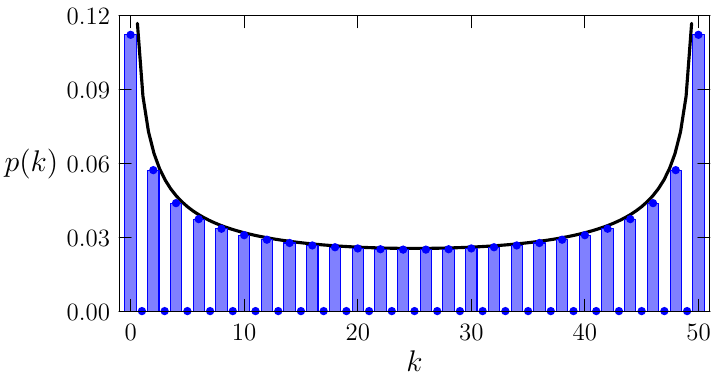}
	\caption{\textbf{Edge states in the generalised non-linear SSH model:} zero-energy weakly localised edge states for a chain with couplings shown in Eq.~(\ref{spin_couplings}) and \textbf{a}, $S+1=11$, \textbf{b}, $S+1=51$ sites. The boundary states are modelled by photon statistics generated by the state in Eq.~(\ref{zero-energy_state}), $p(k)=\bigl\lvert \phi_k^{(1/2)}(0, S) \bigr\rvert^2$. The solid curve is an asymptotic envelope described in the text.}
	\label{edge_states}
\end{figure}

The zero-energy eigenmode for a chain of length $S+1=51$ is shown in Fig.~2a of the main text. Let us find the eigenstate of $H_{\text{BS}}$ that can simulate it. To this end, we will employ the following transformation 
\begin{equation}
	\mathcal{O}^{\dagger }a \, \mathcal{O} = \frac{a-b}{\sqrt{2}}, \qquad \mathcal{O}^{\dagger }b \, \mathcal{O} = \frac{a+b}{\sqrt{2}},
\end{equation} 
where $\mathcal{O} = \exp\{ \tfrac{\pi}{4} (a b^{\dagger} -a^{\dagger}b)\}$, which diagonalises $H_{\text{BS}}$ in the basis of Fock states
\begin{equation}
	H_{\text{BS}}^{\text{diag}} = \mathcal{O}^{\dagger }\, H_{\text{BS}} \, \mathcal{O} = \tfrac{1}{2} \left( a^{\dagger}a - b^{\dagger}b \right).
\end{equation}
Thus, the eigenstates of $H_{\text{BS}}^{\text{diag}}$ are two-mode Fock states
\begin{align}
	H_{\text{BS}}^{\text{diag}} \ket{l,S-l} {}&= (l-\tfrac{S}{2}) \ket{l,S-l}.
\end{align}
From the above we learn about the eigenstates of the original $H_{\text{BS}}$
\begin{align}
	\mathcal{O}^{\dagger } H_{\text{BS}} \mathcal{O} \ket{l,S-l} {}&= (l-\tfrac{S}{2}) \ket{l,S-l},\\
	H_{\text{BS}} \bigl(\mathcal{O} \ket{l,S-l} \bigr) {}&= (l-\tfrac{S}{2})\, \bigl(\mathcal{O}\ket{l,S-l}\bigr).
\end{align}
Thus, the states $\ket{\psi^{(l)}} = \mathcal{O}\ket{l,S-l}$ for $l \in \{0,\dots, S\}$ are the eigenstates of $H_{\text{BS}}$ with corresponding eigenvalues of $l-\frac{S}{2}$. In particular, the eigenstate defined by $l=\tfrac{S}{2}$ corresponds to the zero eigenvalue and thus, simulates the zero-energy mode
\begin{equation}
	H_{\text{BS}} \ket{\psi^{(S/2)}} = H_{\text{BS}} \bigl(\mathcal{O}\ket{\tfrac{S}{2},\tfrac{S}{2}}\bigr) = 0.
\end{equation}

The explicit form of this photonic state is as follows
\begin{align}
	\boxed{\ket{\psi^{(S/2)}} = \mathcal{O}\ket{\tfrac{S}{2},\tfrac{S}{2}} = \sum_{k=0}^S e^{-i \tfrac{\pi}{2} \tfrac{S}{2}}\;
		\phi_k^{\left(1/2\right)}(0, S) \ket{k,S-k}}
	\label{zero-energy_state}
\end{align}
where $\phi_k^{\left(1/2\right)}(0, S)$ are symmetric orthonormal Kravchuk functions. They may be expressed by means of the Gauss hypergeometric function 
\begin{equation}
	\phi_k^{(r)}(l-Sr,S)= (-1)^k \sqrt{\binom{S}{l}\binom{S}{k}} \sqrt{(1-r)^{S-l-k}\,r^{l+k}}
	\, {}_2F_1\left[-k, -l; -S; \tfrac{1}{r}\right].
\end{equation}

The state $\lvert \psi^{(S/2)} \rangle$ differs from the zero-energy eigenmode $\lvert \Psi_0\rangle$ that we simulated and discussed in the main text by a phase factor. Nevertheless its photon statistics, which reads $p(k)=\bigl\lvert \phi_k^{(1/2)}(0, S) \bigr\rvert^2$ and is shown in Fig.~\ref{edge_states}, is identical to these shown in Fig.~2 and 6 in the main text. For large $S$, the function $\frac{4}{\pi S\sqrt{1-(2k/(S-1))^2}}$ provides an asymptotic envelope to $p(k)$, which is indicated as a solid curve in this figure.

\section{Quantum program 2: simulation of perfect quantum state transfer}

\noindent
{\bf Input data initialised to:} $\ket{l=0}_a\ket{S-l=S}_b$.
\smallskip

\noindent
{\bf Program setting:} we run this program for five different settings: $r=0.02$, $0.15$, $0.5$, $0.85$, and $0.98$.

\subsection{The Kravchuk transform}

The $\alpha$-fractional Kravchuk--Fourier transform of an input sequence $\{x_k\}$ for $k=0,1,\ldots,S$ is defined as follows 
\begin{align}
	X_k={}&\sum_{l=0}^S F_{k,l}^{\alpha}\,x_{l},\\
	F_{k,l}^{\alpha}={}& e^{i\tfrac{\pi}{2}(l-k-S\alpha/2)}
	\phi_k^{(p)}(l-Sp,S),
\end{align}
where $\phi_k^{(p)}(l-Sp,S)$ is a Kravchuk function and $p=\sin^2\bigl(\frac{\pi\alpha}{4}\bigr)$.

Mathematically, the beam-splitter interaction in the Fock state basis amounts to an $\alpha$-fractional quantum Kravchuk transform ($\alpha$-QKT) of the input state with fractionality $\alpha = \tfrac{2\theta}{\pi}=\frac{4}{\pi}\arcsin\sqrt{r}$\kern.25em, where $r$ is the beam splitter reflectivity~\cite{Stobinska2019}. In the supplementary material~\cite{Stobinska2019} we have provided analytical computations proving that the probability amplitude of detecting $k$ and $S-k$ photons behind the beam splitter provided that $l$ and $S-l$ were injected into it evaluates the Kravchuk transform
\begin{align}
	\mathcal{A}_S(k,l)={}\bra{k,S-k}U_{\text{BS}}^{(r)}\ket{l,S-l} =
	e^{i\tfrac{\pi}{2}(l-k-S \tfrac{\theta}{\pi})}
	\phi_k^{\left(r\right)}(l-S r, S).
\end{align}

2-QKT is the spatial inversion operator. This becomes clear if we consider the matrix representation of a general beam-splitter interaction 
\begin{align}
	\mathbf{U}_{\text{BS}}={}&\begin{pmatrix}
		\cos\tfrac{\theta}{2} & e^{-i\varphi}\,\sin\tfrac{\theta}{2} \\
		-e^{i\varphi}\,\sin\tfrac{\theta}{2}& \cos\tfrac{\theta}{2} 
	\end{pmatrix},
	\label{rotation}
\end{align}
If we set $r=\sin^2\tfrac{\theta}{2}$
\begin{align}
	\mathbf{U}_{\text{BS}}={}&\begin{pmatrix}
		\sqrt{1-r}& e^{-i\varphi}\sqrt{r}\\
		-e^{i\varphi}\sqrt{r}& \sqrt{1-r}
	\end{pmatrix}.
\end{align}
As $\alpha=2$ corresponds to $r=1$, for $\varphi = \tfrac{\pi}{2}$ we obtain an inversion operation that swaps the input modes
\begin{align}
	\mathbf{U}_{\text{BS}}={}& -i \begin{pmatrix}
		0& 1\\
		1& 0
	\end{pmatrix}.
\end{align}

\subsection{Our simulation: perfect state transfer as a result of mirror reflection}

The beam splitter interaction $U_{\text{BS}}^{(r)}=e^{-i \theta(r) H_{\text{BS}}}$ can simulate the dynamics $U_{\text{XY}}(t)=e^{-i t H_{\text{XY}}}$ of the spin chain (\ref{H_XY}) in its single-excitation subspace. $U_{\text{BS}}^{(1)}$ performs a spatial inversion operation of the sequence $\{x_l\}$, where $l=0,\ldots,S$, for input state $\ket{\Psi}=\sum_{l=0}^S x_l\ket{l}\ket{S-l}$ with respect to the point $l=\frac{S}{2}$\kern.25em~\cite{Hakioglu}. This leads to mapping $\{x_l\}$ to $\{X_l\}=\{x_{S-l}\}$ thus, to the mirror reflection of the input sequence w.r.t.\ the centre of the domain. Since $r=1$ corresponds to $\theta=\pi$, $U_{\text{XY}}(t)$ performs the same operation at $t=\pi$, regardless the input state. 

For any beam-splitter reflectivity, $U_{\text{BS}}^{(r)}$ corresponds to an $\alpha$-QKT which is additive~\cite{Stobinska2019},  $U_{\text{BS}}^{(r(\theta_1+\theta_2))}=U_{\text{BS}}^{(r(\theta_1))}U_{\text{BS}}^{(r(\theta_2))}$, where $r(\theta)=\sin^2\frac{\theta}{2}$. Thus, $U_{\text{BS}}^{(1)}$ can be decomposed into $N$ infinitesimal evolutions $U_{\text{BS}}^{(1)}=\bigotimes_{i=1}^N U_{\text{BS}}^{(r_i)}$, where $\sum_{i=1}^N \theta(r_i)=\theta(1)=\pi$. This property is demonstrated by our quantum simulations performed for subsequent values of $r$, shown in Figs.~2b and 4b in the main text.

\section{Additional simulations: Majorana modes and the Ising model} 

\subsection{Generalised Kitaev model}

The quantum simulations based on Fock-state interference may be reinterpreted in the language of Bogoliubov--de~Gennes Hamiltonians to simulate systems that are not restricted to a single excitation subspace.

A one-dimensional p-wave superconducting chain of $N$ atomic sites is described by the following second quantized Hamiltonian
\begin{equation}
	H_\textrm{K} = \sum_{n=1}^{N}\left\{ -\mu_n (c_n^\dagger c_n - 1/2) -t_n (c_{n+1}^\dagger c_n + c_n^\dagger c_{n+1}) + \Delta_n (c_{n+1}^\dagger c_n^\dagger + c_n c_{n+1}) \right\}.
	\label{kitaevHphoton}
\end{equation}
This is the Kitaev model~\cite{Kitaev2001} but with site dependent chemical potential $\mu_n$, hopping amplitudes $t_n$ and energy gap $\Delta_n$. This Hamiltonian may be diagonalised using the Bogoliubov--de~Gennes trick as follows. First, each term is written in a symmetric form using the fermion anticommutation relations, for example $c_n^\dagger c_n = \frac{1}{2}(c_n^\dagger c_n - c_n c_n^\dagger) + \frac{1}{2}$. Secondly, we introduce the Nambu spinor of all annihilation and creation operators $\chi = (c_1,c_2,\ldots c_1^\dagger,c_2^\dagger,\ldots)^\textrm{T}$. Finally, we write the Hamiltonian as
\begin{equation}
	H_\textrm{K} = \frac{1}{2}\chi^\dagger \mathbf{H}_\textrm{BdG} \chi,
	\label{nambuformphoton}
\end{equation}
where the Bogoliubov--de~Gennes (BdG) Hamiltonian $\mathbf{H}_\textrm{BdG}$ is a $2N\times2N$ matrix that may be interpreted as an effective single particle Hamiltonian for the systems quasiparticles. Notice that the existence of terms like $c_{n+1}^\dagger c_n^\dagger$ and $c_n c_{n+1}$ that don't conserve total particle number force us to include `hole operators' $c_n^\dagger$ in the `vector' $\chi$. This doubles the dimension of our effective single particle Hamiltonian, and leads to the quasiparticles being a combination of particle and hole operators. The BdG Hamiltonian may be directly diagonalised to obtain the energy spectrum of the system.

The second quantized Hamiltonian can equally well be expressed in terms of Majorana operators $a_n = c_n + c_n^\dagger$ and $b_n = i(c_n^\dagger-c_n)$ as
\begin{equation}
	H_\textrm{K} = \frac{i}{2}\sum_{n=1}^{2N}\left\{ -\mu_n a_n b_n + (\Delta_n+t_n) b_n a_{n+1} + (\Delta_n-t_n)a_n b_{n+1}\right\} = \frac{1}{2}\chi'^\dagger \mathbf{H}_\textrm{BdG}' \chi'.
\end{equation}
Here $\chi' = \frac{1}{\sqrt{2}}(a_1, -i b_1, a_2, -i b_2, \ldots a_{N}, -i b_{N})^\textrm{T}$ and
\begin{equation}
	\mathbf{H}_\textrm{BdG}' = 
	\begin{pmatrix}
		0 & \mu_1 & 0 & -\Delta_1 + t_1 & 0 & 0 \\
		\mu_1 & 0 & \Delta_1 + t_1 & 0 & 0 & 0 \\
		0 & \Delta_1 + t_1 & 0 & \mu_2 & 0 & -\Delta_2 + t_2\\
		-\Delta_1 + t_1 & 0 & \mu_2 & 0 & \Delta_2 + t_2 & 0\\
		0 & 0 & 0 & \Delta_2 + t_2 & 0 & \mu_3\\
		0 & 0 & -\Delta_2 + t_2 & 0 & \mu_3 & 0
	\end{pmatrix},
	\label{Hmaj2photon}
\end{equation}
(written for $N=3$ for simplicity) are expressed in the basis of Majorana operators by the following unitary transformation $\chi' = \mathbf{M}\chi$, $\mathbf{H}_\textrm{BdG}' = \mathbf{M} \mathbf{H}_\textrm{BdG} \mathbf{M}^\dagger$ where 
\begin{equation}
	\mathbf{M}
	=
	\frac{1}{\sqrt{2}}
	\begin{pmatrix}
		1 & 0 & 0 & 1 & 0 & 0 \\
		-1 & 0 & 0 & 1 & 0 & 0 \\
		0 & 1 & 0 & 0 & 1 & 0\\
		0 & -1 & 0 & 0 & 1 & 0\\
		0 & 0 & 1 & 0 & 0 & 1\\
		0 & 0 & -1 & 0 & 0 & 1
	\end{pmatrix}.
\end{equation}

\begin{figure}[h]
	\raisebox{1cm}{\textbf{a}}\quad
	\includegraphics[width=9cm]{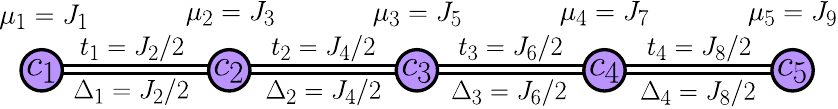}
	\vskip0.5cm
	\raisebox{1cm}{\textbf{b}}\quad
	\includegraphics[width=9cm]{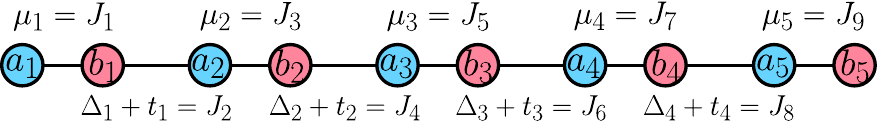}
	\caption{\textbf{Generalised Kitaev chain:} A p-wave superconducting chain with parameters $\mu_n = J_{2n-1}, t_n = \Delta_n = \tfrac{J_{2n}}{2}$ may be simulated by our photonic system. Here the system is depicted in terms of \textbf{a} real Fermion operators and \textbf{b} Majorana operators. In this latter representation, the system shows a strong similarity to the generalised SSH model Fig.~\ref{fig:generalisedSSH}.}
	\label{fig:kitaev}
\end{figure}

If we assign the values $\mu_n = J_{2n-1}, t_n = \Delta_n = \tfrac{J_{2n}}{2}$, where $J_n$ is defined by Eq.~(\ref{spin_couplings}) and $2N = S+1$, $\mathbf{H}_\textrm{BdG}'$ is identical to our non-linear SSH and beam splitter Hamiltonian Eq.~(\ref{hmatrix}). This generalised Kitaev chain of $N$ sites thus has the same quasiparticle energy spectrum as the non-linear SSH chain of $2N$ sites i.e.\ $l-S/2$ for $l \in \{0,\dots, S\}$. The doubling of the energy spectrum is an artefact of the particle-hole symmetry imposed when constructing the Nambu spinor $\Psi$. A single site of the SSH chain is in effect mapped to a single Majorana fermion of the Kitaev chain, as can be seen in Fig.~\ref{fig:kitaev}. The correspondence between the conventional SSH and Kitaev models is well known, being part of the broader equivalence between topological insulators and superconductors~\cite{Cobanera2015}.

\subsection{Interpreting the simulations}

From the quantum simulation we can infer how an initial state evolves in this system. An initial state e.g.\ $c_n^\dagger \ket{\textrm{GS}}$, will evolve after a time $\theta$ into the state $U_\textrm{K}^\dagger c_n^\dagger U_\textrm{K} \ket{\textrm{GS}}$ where $U_\textrm{K} = \exp(-i\theta H_\textrm{K})$ is the unitary operator for the second quantized Hamiltonian $H_\textrm{K}$ and $\ket{\textrm{GS}}$ represents the ground state. Thus by taking combinations of the operators $c_n$ and $c_n^\dagger$ (or equivalently the Majorana operators $a_n$ and $b_n$) we can find the evolution of any state. Alternatively, one may find the evolution of these operators using the smaller BdG matrix $U_\textrm{BdG}(\theta) = \exp(-i\theta H_\textrm{BdG})$. We simply write the operator using the Nambu basis $(c_1,c_2,\ldots c_1^\dagger,c_2^\dagger,\ldots)$ and then operate on it with $U_\textrm{BdG}$. For an example let's take $N=3$ and find the evolution of $c_1^\dagger$. The operator $c_1^\dagger$ is written in the Nambu basis as $(0,0,0,1,0,0)$, its evolution after a time $\theta = \pi$ is then determined by
\begin{equation}
	U_\textrm{BdG}(\pi) 
	\begin{pmatrix}
		0\\
		0\\
		0\\
		1\\
		0\\
		0
	\end{pmatrix}
	=
	\begin{pmatrix}
		0 & 0 & i & 0 & 0 & 0\\
		0 & i & 0 & 0 & 0 & 0\\
		i & 0 & 0 & 0 & 0 & 0\\
		0 & 0 & 0 & 0 & 0 & -i\\
		0 & 0 & 0 & 0 & -i & 0\\
		0 & 0 & 0 & -i & 0 & 0
	\end{pmatrix}
	\begin{pmatrix}
		0\\
		0\\
		0\\
		1\\
		0\\
		0
	\end{pmatrix}
	=
	\begin{pmatrix}
		0\\
		0\\
		0\\
		0\\
		0\\
		-i
	\end{pmatrix}
\end{equation}
indicating that $c_1^\dagger$ evolves into $U_\textrm{K}^\dagger c_1^\dagger U_\textrm{K} = -i c_3^\dagger$ as can be confirmed by direct calculation.

This can also be done using the Majorana basis $\frac{1}{\sqrt{2}}(a_1, -ib_1,\ldots a_N, -ib_N)$ and $U_\textrm{BdG}'(\theta) = \exp(-i\theta H_\textrm{BdG}')$. The same calculation as above can then be performed, the operator $c_1^\dagger = \frac{1}{2}(a_1 - ib_1)$ is now written as $(1/\sqrt{2},1/\sqrt{2},0,0,0,0)$ and we evaluate
\begin{equation}
	U_\textrm{BdG}'(\pi) 
	\begin{pmatrix}
		1/\sqrt{2}\\
		1/\sqrt{2}\\
		0\\
		0\\
		0\\
		0
	\end{pmatrix}
	=
	\begin{pmatrix}
		0 & 0 & 0 & 0 & 0 & -i\\
		0 & 0 & 0 & 0 & -i & 0\\
		0 & 0 & 0 & -i & 0 & 0\\
		0 & 0 & -i & 0 & 0 & 0\\
		0 & -i & 0 & 0 & 0 & 0\\
		-i & 0 & 0 & 0 & 0 & 0
	\end{pmatrix}
	\begin{pmatrix}
		1/\sqrt{2}\\
		1/\sqrt{2}\\
		0\\
		0\\
		0\\
		0
	\end{pmatrix}
	=
	\begin{pmatrix}
		0\\
		0\\
		0\\
		0\\
		-i/\sqrt{2}\\
		-i/\sqrt{2}
	\end{pmatrix}
\end{equation}
indicating that $c_1^\dagger = \frac{1}{2}(a_1 - ib_1)$ evolves into $\frac{-i}{2}(a_3-ib_3) = -ic_3^\dagger$ as before.

For our chosen parameters $U_\textrm{BdG}'(\theta)$ is identical to the beam splitter unitary operator where $\theta = 2\arcsin(\sqrt{r})$ and $r$ is the beam splitter reflectivity. The above calculation then tells us how to perform the photonic simulation. To simulate a p-wave chain of $N$ sites we take a photonic system of $S = 2N-1$ photons and associate the two mode Fock states with the Majorana operators as 
\begin{equation}
	\ket{2(n-1)}_a\ket{S-2(n-1)}_b \leftrightarrow \frac{a_n}{\sqrt{2}},\quad \ket{2n-1}_a\ket{S-(2n-1)}_b \leftrightarrow \frac{-ib_n}{\sqrt{2}}.
	\label{labels}
\end{equation}
To find how the operators evolves after a time $\theta$ we interfere the corresponding two mode Fock state on a beam splitter of reflectivity $r = \sin^2(\theta/2)$. To find how the real fermion operators transform, e.g.\ $c_1^\dagger = \frac{1}{2}(a_1 - ib_1)$, we can take superpositions $U_\textrm{BS}\frac{1}{\sqrt{2}} (\ket{0}_a\ket{S}_b + \ket{1}_a\ket{S-1}_b)$. In a similar way, the evolution of multiply excited states e.g.\ $U_\textrm{K}^\dagger c_1^\dagger c_2^\dagger U_\textrm{K}$ may be determined by decomposing into $(U_\textrm{K}^\dagger c_1^\dagger U_\textrm{K})( U_\textrm{K}^\dagger c_2^\dagger U_\textrm{K})$ and finding the evolution of the bracketed terms individually, which determines the evolution of the final state up to the global phase. One note of caution is that if we just perform photon number measurements we cannot distinguish between the states $\frac{1}{\sqrt{2}}(\ket{S-1}_a\ket{1}_b \pm \ket{S}_a\ket{0}_b)$ and thus don't know if we obtain $c_3 = \frac{1}{2}(a_3 + ib_3)$ or $c_3^\dagger = \frac{1}{2}(a_3 - ib_3)$. This problem may be easily resolved by extending the experimental detection to obtain more tomographically complete information, using e.g.\ homodyne detection.

\subsubsection*{Perfect transfer of Majorana fermions}

Just like the generalised SSH model, this new system allows the perfect transfer of a quantum state due to its correspondence with the beam-splitter dynamics. After an interaction time $\theta = \pi$ an initial state localised at one end of the chain $c_1^\dagger \ket{\textrm{GS}}$ is perfectly transferred to the other end $U_\textrm{K}^\dagger c_1^\dagger U_\textrm{K}\ket{\textrm{GS}} = -i c_N^\dagger \ket{\textrm{GS}}$. This transfer can be viewed in terms of Majorana operators, in particular the two Majoranas at the ends of the chain swap after the interaction time $\theta = \pi$
\begin{equation}
	a_1 \to -b_{N}, \quad b_{N} \to a_1,
\end{equation}
reminiscent of a Majorana braiding operation~\cite{Beenakker2013}. If the system is left to evolve further these operators will keep evolving as, unlike the original Kitaev model, they are not zero energy modes. However, the above transfer could be viewed as an intermediate process, with the parameters being engineered from the original Kitaev values, $\mu_n = 0$, $t_n = \Delta_n = \textrm{constant}$, to the above values only during the transfer process. The physical engineering of such a system is extremely difficult in a condensed matter setting, but is possible in systems of 1-dimensional photonic cavities where the effective Majorana modes emerge due to Kerr-type non-linearities within a Bose--Hubbard model~\cite{Bardyn2012}.

\subsection{Transverse-field Ising Model}

The Hamiltonian Eq.~(\ref{kitaevHphoton}) may be mapped to a spin chain using the Jordan--Wigner transformation
\begin{equation}
	c_n = \prod_{m=1}^{n-1} (-\sigma_m^z) \sigma_n^+, \quad
	c_n^\dagger = \prod_{m=1}^{n-1} (-\sigma_m^z) \sigma_n^-, \label{jw}
\end{equation}
where $\sigma_m^z$ acts on the $m$th spin of the chain and similarly for the other Pauli operators. The resulting Hamiltonian is
\begin{equation}
	H_\textrm{K} = \frac{1}{2}\sum_{n=1}^N\{\mu_n \sigma_n^z + (t_n+\Delta_n)\sigma_n^x \sigma_{n+1}^x + (t_n-\Delta_n)\sigma_n^y \sigma_{n+1}^y\}.
\end{equation}
For the couplings we consider $\mu_n = J_{2n-1}, t_n = \Delta_n = \tfrac{J_{2n}}{2}$, this becomes
\begin{equation}
	H_\textrm{K} = \frac{1}{2}\sum_{n=1}^N\{\mu_n \sigma_n^z + 2t_n \sigma_n^x \sigma_{n+1}^x\},
\end{equation}
which is an Ising chain in a non-uniform transverse field. Thus in principle we can simulate this system using our Fock-state interference platform. One inverts the transformations Eq.~(\ref{jw}) and then uses the correspondence Eq.~(\ref{labels}) to associate the two-mode Fock states with combinations of spin operators. For a concrete example we again take $N=3$ $(S=2N-1=5)$, the Jordan--Wigner transformation is then
\begin{align}
	c_1 = \sigma_1^+,\quad c_2 = -\sigma_1^z \sigma_2^+,\quad c_3 = \sigma_1^z \sigma_2^z \sigma_3^+,\\
	c_1^\dagger = \sigma_1^-,\quad c_2^\dagger = -\sigma_1^z \sigma_2^-,\quad c_3^\dagger = \sigma_1^z \sigma_2^z \sigma_3^-,\nonumber
\end{align}
while, using Eq.~(\ref{labels}) and $c_n = \frac{1}{2}(a_n + i b_n)$, the mapping to Fock states is
\begin{align}
	\frac{1}{\sqrt{2}}(\ket{0}_a\ket{5}_b - \ket{1}_a\ket{4}_b) \leftrightarrow \sigma_1^+,\quad \frac{1}{\sqrt{2}}(\ket{2}_a\ket{3}_b - \ket{3}_a\ket{2}_b) \leftrightarrow -\sigma_1^z \sigma_2^+,\quad \frac{1}{\sqrt{2}}(\ket{4}_a\ket{1}_b - \ket{5}_a\ket{0}_b) \leftrightarrow \sigma_1^z \sigma_2^z \sigma_3^+,\label{twomodecorr}\\
	\frac{1}{\sqrt{2}}(\ket{0}_a\ket{5}_b + \ket{1}_a\ket{4}_b) \leftrightarrow \sigma_1^-,\quad \frac{1}{\sqrt{2}}(\ket{2}_a\ket{3}_b + \ket{3}_a\ket{2}_b) \leftrightarrow -\sigma_1^z \sigma_2^-,\quad \frac{1}{\sqrt{2}}(\ket{4}_a\ket{1}_b + \ket{5}_a\ket{0}_b) \leftrightarrow \sigma_1^z \sigma_2^z \sigma_3^-\nonumber.
\end{align}

To simulate the evolution of a spin initially at the first site, we consider the initial state $\sigma_{1}^+ \ket{\downarrow_1, \downarrow_2, \downarrow_3}$. The final state after a time $\theta(r)$ is then found from interfering the two-mode state $\frac{1}{\sqrt{2}}(\ket{0}_a\ket{5}_b - \ket{1}_a\ket{4}_b)$ on a beam splitter of reflectivity $r$. Taking again $\theta = \pi$ $(r = 1)$ as an example, we know that the perfect reflectivity implies that output state is $\frac{1}{\sqrt{2}}(\ket{4}_a\ket{1}_b - \ket{5}_a\ket{0}_b)$. Using the table Eq.~(\ref{twomodecorr}), we thus conclude that the final state of the spin system is $\sigma_1^z \sigma_2^z \sigma_3^+ \ket{\downarrow_1, \downarrow_2, \downarrow_3}$. The leading operators $\sigma_n^z = 1 - 2\sigma_n^-\sigma_n^+$ only provide a phase factor and do not flip any spins so that the initial spin is transferred to the other end of the chain, just as we saw in section IV. The Ising model with uniform parameters has been previously studied in the context of quantum state transfer~\cite{Yao2013}, our simulations suggest that modifying the parameters as described above could improve the situation, as the system inherits the perfect reflection property of the beam splitter when $\theta = \pi$.

\subsection{Relationship between simulated systems}

\begin{figure}[h]
	\includegraphics[width=6cm]{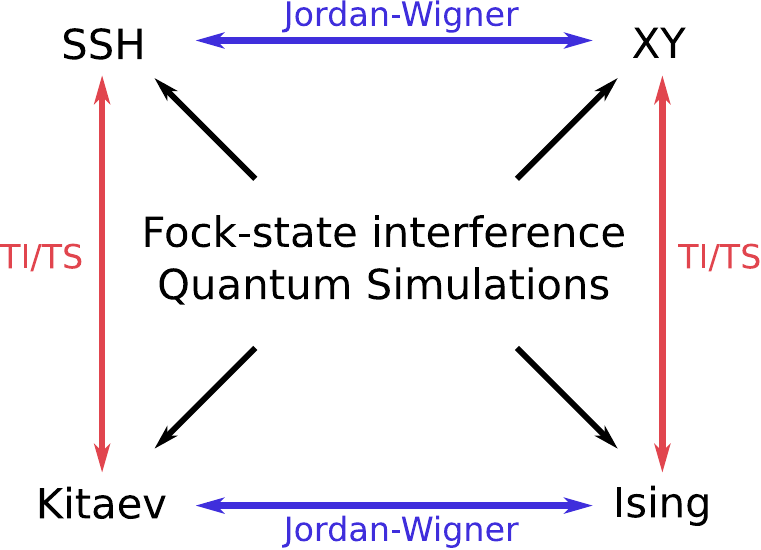}
	\caption{\textbf{Relationship between simulated systems:} The four example systems that we have discussed are interrelated by both Jordan--Wigner transformations and the topological insulator/ topological superconductor correspondence.}
	\label{fig:4systems}
\end{figure}

We have discussed four example systems that may be simulated by our platform. The XY spin chain and non-linear SSH model are, strictly speaking, related by a Jordan--Wigner transformation, with the latter being expressed in terms of Fermionic operators in a crystal lattice. However, due to the conservation of total spin / particle number we can restrict ourselves to the single excitation subspace whereupon these two systems have identical interpretations and the terms are used interchangeably in the text. The other two systems, the non-uniform Ising and Kitaev models, are obtained from the XY spin chain and SSH models by a correspondence between topological insulating and topological superconducting systems. In simpler terms, we map the beam splitter Hamiltonian to Bogoliubov--de~Gennes Hamiltonians rather than single particle ones, which gives the simulations a slightly more complicated interpretation. We have depicted the relationships between these simulated systems in Fig.~\ref{fig:4systems} for clarity. 

\section{Extension to multiple dimensions}

\begin{figure}[h]
	\includegraphics[width=6cm]{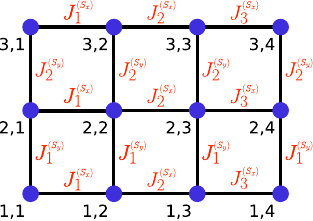}
	\caption{\textbf{2-dimensional XY model:} A lattice of size $(S_x+1) \times (S_y+1)$ may be simulated by taking the four-mode Fock states $\ket{i}_a\ket{S_x-i}_b \ket{j}_{a'}\ket{S_y-j}_{b'}$ and interfering the modes $a,b$ and $a',b'$ together. The system is here depicted for $S_x = 3$, $S_y = 2$.}
	\label{fig:2d}
\end{figure}

As an example of how to extend our simulations to multiple dimensions, we take the XY spin model (or equivalently, generalised SSH model) in a two dimensional rectangular lattice. The Hamiltonian for a lattice of size $(S_x+1) \times (S_y+1)$ is
\begin{equation}
	H_{\text{XY}}^{(2)} ={} \sum_{m=1}^{S_y+1} \sum_{n=1}^{S_x} J_n^{(S_x)} \left( \sigma_{m,n}^+ \sigma_{m,n+1}^- + \textrm{h.c.}\right) + \sum_{n=1}^{S_x+1}\sum_{m=1}^{S_y} J_m^{(S_y)} \left( \sigma_{m,n}^+ \sigma_{m+1,n}^- + \textrm{h.c.}\right).
	\label{H_XY2}
\end{equation}
Here the operators $\sigma_{m,n}^+$ etc. act on the spin at site $(m,n)$, the terms $J_n^{(S_x)}$ denote couplings along the x-axis of the lattice, while $J_n^{(S_y)}$ denote couplings along the y-axis. When $S_y = 0$, this reduces to the 1-dimensional case Eq.~(\ref{H_XY}). The system is depicted in Fig.~(\ref{fig:2d}). Just as in the 1-dimensional case we look at the Hamiltonian matrix elements in the single excitation subspace. Denoting the state $\ket{i,j} = \sigma_{i,j}^+ \ket{\downarrow_{1,1} \ldots \downarrow_{S_x+1,S_y+1}}$, these are
\begin{equation}
	\bra{p,q}H_{XY}^{(2)}\ket{i,j} = \delta_{p,i} (J_q^{(S_x)} \delta_{q,j-1} + J_j^{(S_x)} \delta_{q,j+1}) + \delta_{q,j}(J_p^{(S_y)} \delta_{p,i-1} + J_i^{(S_y)} \delta_{p,i+1}).
	\label{2dmatrixelements}
\end{equation}

To perform the simulation, we take four photonic modes which we label $a,a',b,b'$. The modes $a$ and $b$ are interfered on a beam splitter, and similarly with $a'$ and $b'$. This could be achieved with two different beam splitters, or the same beam splitter in different frequency, polarisation or spatio-temporal modes. The total Hamiltonian is $H_{BS} = \frac{1}{2}(a^\dagger b + ab^\dagger + a'^\dagger b' + a' b'^\dagger)$ and in the Fock state basis $\ket{i}_a\ket{S_x-i}_b \ket{j}_{a'}\ket{S_y-j}_{b'}$ one can check that the matrix elements are identical to Eq.~(\ref{2dmatrixelements}) provided that $J_n^{(S)} = \tfrac{1}{2}\sqrt{n(S+1-n)}$. To simulate a system of size $(S_x+1) \times (S_y+1)$ one simply interferes the corresponding Fock states, with $S_x$ total photons in modes $a$ and $b$, and $S_y$ total photons in modes $a'$ and $b'$.

\clearpage\twocolumngrid

\end{document}